\documentclass[prd,twocolumn,preprintnumbers,amsmath,amssymb,nofootinbib,epsfig,bmamsfonts,yfonts,superscriptaddress]{revtex4}
\usepackage{graphicx}
\usepackage{lipsum}
\usepackage{mathtools}
\usepackage{hyperref}

\def \cf {C_F}
\def \nc {N_c}
\def \ca {C_A}
\def \nf {n_f}
\def \tf {T_F}

\newcommand{\Tint}[1]{{\hbox{$\sum$}\!\!\!\!\!\!\!\int\,}_{\!\!\!\!\raise-0.9ex\hbox{$\scriptstyle{#1}$}}}

\def\siml{{\ \lower-1.2pt\vbox{\hbox{\rlap{$<$}\lower6pt\vbox{\hbox{$\sim$}}}}\ }}
\def\simg{{\ \lower-1.2pt\vbox{\hbox{\rlap{$>$}\lower6pt\vbox{\hbox{$\sim$}}}}\ }}

\def \als {\alpha_{\mathrm{s}}}

\def \m2   {\mu^{2 \epsilon}}

\def\siml{{\ \lower-1.2pt\vbox{\hbox{\rlap{$<$}\lower6pt\vbox{\hbox{$\sim$}}}}\ }}
\def\simg{{\ \lower-1.2pt\vbox{\hbox{\rlap{$>$}\lower6pt\vbox{\hbox{$\sim$}}}}\ }}

\def\nn {\nonumber}

\def \OO {\mathcal{O}}

\relax

\begin{document}

\title{Thermal width of the Higgs boson in hot QCD matter}

\author{Jacopo Ghiglieri}
\email{jacopo.ghiglieri@cern.ch}
\affiliation{Theoretical Physics Department, CERN, CH-1211 Gen\`eve 23, Switzerland}
\author{Urs Achim Wiedemann}
\email{urs.wiedemann@cern.ch}
\affiliation{Theoretical Physics Department, CERN, CH-1211 Gen\`eve 23, Switzerland}

\preprint{CERN-TH-2019-003}

\begin{abstract}
Following Caron-Huot  and combining results for the thermal dependence of spectral functions at large time-like momenta, we write an explicit expression for the thermal width of the Higgs boson to $\OO(\als)$
for $T \ll M_H$. It is an $ \OO\left( \als \left( \textstyle\frac{T}{M_H}\right)^4 \right)$ correction for $H\to gg$ and $H\to q\bar{q}$.   We also compile corresponding results for the thermal width of the $Z$-boson, and we recall which generic structures of the field theory, accessible via the operator product expansion, fix the $\frac{T}{M}$-dependence of the decay of heavy particles.
\end{abstract}

\maketitle

\section{Introduction}
In this paper we derive the thermal correction $\delta\Gamma_{H}$ of the width of the Higgs boson
to first order in the strong coupling constant $\als$ and for temperatures $T$ of the QCD plasma that are
parametrically lower than the Higgs mass, $M_H \gg T$. We do so since we could not find an explicit expression
for  $\delta\Gamma_{H}$ in the literature when discussing the (im)possibility of observing thermal
corrections to Higgs branching ratios at future multi-TeV heavy ion collider experiments~\cite{Berger:2018mtg,Citron:2018lsq,dEnterria:2018bqi}.
As we explain below, $\delta\Gamma_{H}$ can be obtained essentially from combining limiting cases of
several spectral functions whose derivation has been described in detail~\cite{CaronHuot:2009ns,Meyer:2010ii,Laine:2010tc}.
We believe this to be known to
a small group of experts in thermal field theory, and the novelty of the present work thus resides mainly in making this
expert knowledge explicit.

For thermal corrections to the Higgs width in a QCD plasma, the branching into final states without color charge
(such as $H\to Z Z\to 4\, l$) is clearly unimportant. The decay processes relevant for the following are therefore
determined by the electroweak interaction of the Higgs to quarks,
\begin{equation}
	\label{eq1}
\mathcal{L}_{Hq}=- S \frac{H}{v}\, ,\qquad S\equiv m_q\bar\psi_q\psi_q\, ,
\end{equation}
and by the corresponding coupling of the Higgs to gluons.
Here, $v\cong 246$ GeV denotes the Higgs vacuum expectation value.
The dominant contribution to $H \to g g$  proceeds
via a top quark loop. We work in the limit $m_t \gg M_H$ in which this interaction is given by the Higgs effective field theory
Lagrangian~\cite{Inami:1982xt}
\begin{eqnarray}
	\label{eq2}
	\mathcal{L}^\mathrm{eff}_{Hg} &=& -C_{Hg}\frac{H}{v}\mathcal{O}_{Hg}\, ,\\
	\mathcal{O}_{Hg} &\equiv& -\frac{1}{4}F^a_{\mu\nu}F^{a\,\mu\nu}\, ,\\
	C_{Hg}&=& \frac{\als}{3\pi}+\OO(\als^2)\, .
\end{eqnarray}

For a particle that does not carry charges of the plasma and that couples to currents $J$, the decay widths can be expressed in terms of the
corresponding spectral functions
\begin{equation}
\label{eq3}
\rho_J(K)\equiv \int d^4x\, e^{i K\cdot X}\,\langle[J(X),J(0)]\rangle\, ,
\end{equation}
where $K_\mu = (k_0, \vec{k})$ in the rest frame of the QCD plasma. The metric is mostly minus.
In particular, the partial decay widths of the Higgs boson relevant for our study are given by
\begin{eqnarray}
	\label{eq4}
		\Gamma_{H\to q\bar q}&=&\frac{1}{v^2}\frac{1}{2k^0}\rho_{S}(K)\, , \\
	\label{eq5}
		\Gamma_{H\to gg}&=& \frac{\als^2}{(3\pi)^2v^2}\frac{1}{2k^0}\rho_{\mathcal{O}_{Hg}}(K)\, ,
\end{eqnarray}
where $\textstyle\frac{1}{2k_0}$ is the usual kinematical flux factor and the factors $\textstyle\frac{1}{v^2}$,
$\textstyle\frac{\als^2}{(3\pi)^2\, v^2}$ denote the squares of the couplings of the Higgs boson to the corresponding currents.
The widths (\ref{eq4}), (\ref{eq5}) are thus first order in the electroweak couplings, but the spectral functions $\rho_J$ are all
orders in $\als$.

The calculation of thermal corrections to $\Gamma_{H\to q\bar q}$, $\Gamma_{H\to gg}$ then amounts to determining thermal
corrections $\delta \rho_{S}(K)$ and $\delta\rho_{\mathcal{O}_{Hg}}(K)$ to the vacuum spectral functions
$\rho_{S}^\mathrm{vac}(K^2)$ and $\rho^\mathrm{vac}_{\mathcal{O}_{Hg}}(K^2)$. Lorentz invariance of the vacuum implies
that the latter can depend only on the scalar $K^2 = K_\mu\, K^\mu$. In contrast, any finite temperature system singles out a
rest frame, and the thermal corrections $\delta \rho_{S}$ and $\delta\rho_{\mathcal{O}_{Hg}}$ can therefore depend separately on
$k_0$ and $k \equiv \vert \vec{k}\vert$. In the present paper, we focus on the case of a Higgs boson at rest in the plasma,
except for a short discussion of boosted Higgs bosons in section~\ref{sec2.3}.

In a QCD plasma, the vacuum branchings $H\to g\, g$ and $H\to q\, \bar{q}$  are modified already to zeroth order in $\als$, since
emission of each final state gluon or final state quark of momentum $k_{g/q}$ is enhanced by a thermal Bose-Einstein
$\left( 1 + f_B(k_g)\right)$ or suppressed by a Fermi-Dirac $\left( 1 - f_F(k_q)\right)$ distribution factor, respectively. However, the partons
emerging from this two-body decay carry momenta $k_{q/g} = \textstyle\frac{M_H}{2}$ much above the thermal scale.
As a consequence, the effects of stimulated emission for the decay into gluons and of Pauli-blocking for the decay into quarks
are negligible. To zeroth order in $\als$,
thermal corrections to $\Gamma_{H\to gg}$ and $\Gamma_{H\to q\bar q}$ are kinematically suppressed by multiplicative factors
$\exp\left[-\textstyle\frac{k_{g/q}}{T}\right] = \exp\left[-\textstyle\frac{M_H}{2T}\right]$.

 Processes to first order in $\als$ open up a region of phase space in which thermal corrections are not suppressed in this way.
Firstly, to $O(\als)$, there are {\it real emission contributions}, such as $H\to g\, g\, g$ or $H\to q\, \bar{q}\, g$. In these
 three-body decays, one of the three final
state partons can carry a momentum $k_{g/q} \lesssim T$, for which effects of stimulated gluon emission
$\propto \left( 1 + f_B(k_g)\right)$ and Pauli-blocked quark emission $\propto \left( 1 - f_F(k_q)\right)$ are not suppressed.
Secondly, there are {\it real absorption contributions} in which quarks and gluons from the QCD plasma interact with the vacuum
branching process, such as $g\, H\to g\, g$, $g\, H\to q\, \bar{q}$, $q\, H\to q\, g$ or $\bar{q}\, H\to g\, \bar{q}$. Thirdly,
there are thermal {\it virtual contribution} that arise from branching processes which interact on the
amplitude level to $O(\als)$ with partons in the medium, and which interfere with the vacuum contribution in the
complex conjugate amplitude. In general, calculations of thermal widths to $O(\als)$
amount to determining these three classes of contributions consistently in finite temperature field theory.
There are two conceptually different approaches for achieving this:

\begin{enumerate}
	\item {\it Explicit perturbative calculation of $\delta\Gamma$ in finite temperature QCD.}\\
	This standard approach is well documented e.g. for calculations of the thermal production
	of vector bosons~\cite{Baier:1988xv,Altherr:1989jc,Altherr:1988bg,Gabellini:1989yk,Kapusta:2000ct,Majumder:2001iy}.
	It is typically formulated in terms of
	the two-loop self-energy correction of the propagator of the particle
	excitation in whose width one is interested.
	By the optical theorem, the imaginary part of this self energy corresponds to a sum over different cut contributions that can be
	identified with the three above-mentioned classes of thermal corrections, namely real emission, real absorption and virtual
	correction. In practice, all three classes of thermal corrections yield infrared singular expressions, while thermal
	corrections to the sum of the three contributions are infrared and collinear safe observables~\cite{Bellac:2011kqa}.
	This makes explicit perturbative calculations of $\delta \Gamma$ relatively complex and lengthy.
	\item {\it Calculating $\delta\Gamma$ via the operator product expansion (OPE) of the relevant spectral functions.}\\
	In general, the OPE relies on a systematic separation of infrared and ultraviolet contributions. As first pointed out
	by Caron-Huot~\cite{CaronHuot:2009ns}, this approach allows one to determine thermal
	corrections to spectral functions in the high-energy time-like region $k_0 \gg T$. For the thermal width of particles
	whose mass is parametrically larger than the plasma temperature, this allows for a much simplified calculation.
\end{enumerate}

A particularly simple and instructive example is the case of a heavy fermion decaying to a lighter fermion and a scalar in a QED plasma.
This process was studied in an explicit perturbative calculation in Ref.~\cite{Czarnecki:2011mr}, presenting explicit IR-regulated results for the real emission, real absorption
and virtual correction contributions, and demonstrating the IR-finiteness of the physical width in detail. The same process was studied in Ref.~\cite{Beneke:2016ghp} with
OPE techniques. (As the decaying particle in this toy model is charged under the gauge group, the techniques of Ref.~\cite{CaronHuot:2009ns} do not apply directly to this case.)

OPE techniques have been applied also to study the thermal corrections to the decay width of hypothetical heavy right-handed neutrinos in~\cite{Laine:2011pq}
(see also~\cite{Salvio:2011sf} for the same calculation with the explicit method). Furthermore, in the case of heavy right-handed neutrinos an Effective Field Theory approach has been introduced
in~\cite{Biondini:2013xua}, where the $M\gg T$ expansion is introduced at the Lagrangian level, making the separation of IR and UV extremely transparent (see~\cite{Biondini:2017rpb} for a review of these calculations in their physical context).

In the present work, we utilize the OPE approach of Ref.~\cite{CaronHuot:2009ns}  to arrive at an expression for the thermal width of the Higgs boson.
In appendix~\ref{appa}, we comment shortly on how these results are connected to results obtained from an explicit perturbative calculation.
In appendix~\ref{appb}, we also summarize results for the thermal width of the $Z$-boson.

\section{Higgs branching ratios from known spectral functions of $T^{\mu\nu}$ }

For the reader who wants to get to the final result for the thermal width $\delta \Gamma_H$ without spending
too much time on technical details, we compile in this section what is known about the spectral functions
$\rho_{\mathcal{O}_{Hg}}(K)$ and $\rho_{S}(K)$ in the asymptotic limit of large $K^2$, and we insert this
information for $K^2 = M_H^2$ into eqs.~(\ref{eq4}) and (\ref{eq5}) to obtain $\delta \Gamma_H$.
 A more thorough discussion of the theoretical basis of this schematic derivation
  is deferred to the subsequent section.

\subsection{$H\to g\, g$}

To determine $\rho_{\mathcal{O}_{Hg}}(K)$, we exploit that the operator $\mathcal{O}_{Hg}$ appears in the trace of
the QCD energy-momentum tensor~\cite{CaronHuot:2009ns}
\begin{equation}
T^{\mu}_{\, \, \mu} = \frac{-b_0\, \als}{2\pi} \mathcal{O}_{Hg} + \hbox{[fermionic terms]}\, ,
\label{eq6}
\end{equation}
where $b_0= (\frac{11}{3}\ca {-} \frac43\nf\tf)$ is the leading coefficient of the $\beta$-function,
with $\ca=\nc=3$ the quadratic Casimir of the adjoint
representation  and $\tf=\frac12$.
The spectral function of the trace anomaly (\ref{eq6}) is the spectral function $\rho_\zeta$ of the bulk viscous channel of $T^{\mu\nu}$.
In general, due to the fermionic mass contributions, $\rho_\zeta$ differs from $\rho_{\mathcal{O}_{Hg}}$ not only by a trivial
prefactor $\textstyle \frac{b_0^2\, \als^2}{(2\pi)^2}$, but also by additional fermionic terms.
However, in the OPE of  $\rho_\zeta$, these fermionic terms are suppressed by
additional powers of $m_q^2/K^2$ which render them negligible for our problem.
From the result for  $\rho_{\zeta}^\mathrm{vac}(K^2)$ in Ref.~\cite{CaronHuot:2009ns}, one thus finds
\begin{eqnarray}
	\rho_{\mathcal{O}_{Hg}}^\mathrm{vac}(K^2) &=& \frac{(2\pi)^2}{b_0^2\, \als^2}\,
	\left( \rho_{\zeta}^\mathrm{vac}(K^2) + \OO\left(\frac{m_q^2}{K^2}\right) \right) \nonumber \\
	&=& \frac{ d_A (K^2)^2}{32 \pi}+\OO(\als)+ \OO\left(\frac{m_q^2}{K^2}\right)\, ,
	\label{eq7}
\end{eqnarray}
where $d_A=N_c^2-1$ is the dimension of the adjoint representation.

The leading (dimension-four) thermal correction to $\rho_{\zeta}$ was derived in the same Ref.~\cite{CaronHuot:2009ns}
up to an unknown coefficient in front of the trace anomaly that has been determined in Ref.~\cite{Meyer:2010ii,Laine:2010tc}.
Accounting again for the fact that
$\delta\rho_{\zeta}(K)$ and $\delta\rho_{\mathcal{O}_{Hg}}(K)$ differ by the prefactor $\textstyle \frac{b_0^2\, \als^2}{(2\pi)^2}$,
these results translate into
\begin{eqnarray}
	\delta \rho_{\OO_{Hg}}(K)  &=&
	  \frac{ 2\als}{3} \frac{K_\mu K_\nu}{K^2}
	      \left[2\cf T^{\mu\nu}_f -
	  (\nf \tf{+}\frac32 b_0) T^{\mu\nu}_g \right] \nonumber \\
	  &&	- \pi T^\mu{}_\mu\, ,
	  \label{eq8}
\end{eqnarray}
where $T^{\mu\nu}_g$ and $T^{\mu\nu}_f$  denote the traceless parts of the gluonic and fermionic contributions to $T^{\mu\nu}$,
respectively, and $\cf=(\nc^2-1)/(2\nc)$ is the quadratic Casimir of the fundamental
representation.  The temperature dependence of $\delta \rho_{\OO_{Hg}}(K)$ thus enters via the temperature dependence of the
energy-momentum tensor.

\subsubsection{Vacuum width from spectral function}
\label{sec2.1.1}
Before turning to a discussion of the thermal corrections, we check the consistency of our starting point by determining
the corresponding partial width in vacuum. Plugging eq.~\eqref{eq7} into \eqref{eq5}, we find
\begin{equation}
		\Gamma^\mathrm{vac}_{H\to gg}
		=\frac{\als^2M_H^3}{72\pi^3 v^2}+\OO(\als^3)\, ,
		\label{eq9}
\end{equation}
which agrees with the expression in the literature~\cite{Inami:1982xt}.

This argument can be extended to next-to-leading order (NLO).
Higher order corrections to the Wilson coefficient $C_{Hg}$ can be found in
\cite{Inami:1982xt,Kramer:1996iq,Chetyrkin:1997un,Steinhauser:2002rq,Grigo:2014jma}

\begin{equation}
	C_{Hg}=\frac{\als}{3\pi} \Bigg\lbrace 1 + \left(\frac{5}{4}C_A - \frac{3}{4}C_F \right)\frac{\als}{\pi}  \Bigg\rbrace\, .
		\label{eq10}
\end{equation}
The vacuum pure glue part of the bulk channel spectral function $\rho_\zeta$
at NLO can be found e.g. in Ref.~\cite{Laine:2011xm}. Multiplying this with the prefactor $\textstyle \frac{b_0^2\, \als^2}{(2\pi)^2}$
of eq.~(\ref{eq6}) yields
\begin{equation}
	\rho_{\mathcal{O}_{Hg}}^\mathrm{vac, NLO}(K^2) = \rho_{\mathcal{O}_{Hg}}^\mathrm{vac}(K^2)
		\left(1 + \frac{\als N_c}{4\pi}\frac{73}{3} \right)\, .
		\label{eq11}
\end{equation}
Combining these expressions, one finds for the NLO correction to $ \Gamma^\mathrm{vac}_{H\to gg}$
 \begin{eqnarray}
  &&\!\!\! \Gamma^\mathrm{vac, NLO}_{H\to gg(g)\, , H\to gq\bar{q}}    \label{eq12} \\
  &&\!\!\!=    \Gamma^\mathrm{vac}_{H\to gg}
  \Bigg\lbrace  1 + \frac{\als}{4\pi}  \left( 2 \left( 5\, N_c - 3\, C_F \right) +  N_c \frac{73}{3}  \right)  \Bigg\rbrace\, .  \nonumber
 \end{eqnarray}
 This is consistent with the NLO correction factor for the pure glue part, $\left( 1 + \textstyle\frac{95}{4} \textstyle\frac{\als}{\pi} \right)$~\cite{Inami:1982xt,Djouadi:1991tka}.
 Eq.~\eqref{eq11} contains to $\OO(\als)$ also a logarithmic term that can be traced back to the RG evolution of the LO result and that is consistent with ~\cite{Inami:1982xt,Djouadi:1991tka}.
 Indeed, we are following here essentially the logic of Ref.~\cite{Inami:1982xt}. Our reason for repeating this result is that we take in the following thermal corrections to
 (\ref{eq11}) from published results in which also the vacuum contribution to the spectral function is given. The rederivation of (\ref{eq9}) and  (\ref{eq12}) thus serves as  a check
 that these thermal corrections are used with proper normalization.

\subsubsection{Thermal corrections to $H\to g g$}

Paralleling the discussion in section~\ref{sec2.1.1},
we obtain the thermal correction to $\Gamma^\mathrm{vac}_{H\to gg}$
from the thermal contribution to the quark and gluon condensates that appear in the OPE of the bulk channel
spectral function eq.~(\ref{eq8}),
\begin{eqnarray}
		\delta\Gamma_{H\to gg} &=& \frac{\als^3}{81\pi^2v^2k^0} \frac{3k_0^2+k^2}{M_H^2}
	      \left[2\cf \bigg(\sum_{q\in udsc}\langle T^{00}_{fq}\rangle \bigg) \right. \nonumber \\
	&& \left.	      -
	  \bigg( \frac{11}{2}\ca - \nf \tf \bigg)\langle  T^{00}_g\rangle  \right]+\OO(\als^4)\, .
	  \label{eq13}
\end{eqnarray}
Here, we have used that in an isotropic medium, the traceless operators $T^{\mu\nu}_{f,g}$ satisfy
$T^{ij}_{f,g} = \textstyle\frac{1}{3} \delta^{ij}\, T^{00}_{f,g}$ and $T^{oj}_{f,g} = 0$. The resulting prefactor
$(3k_0^2+k^2)$ breaks Lorentz invariance since the QCD plasma specifies a thermal rest frame.

In close analogy to the NLO vacuum correction (\ref{eq12}) to $ \Gamma^\mathrm{vac}_{H\to gg}$, also the
$\OO(\als)$ thermal correction $\delta\Gamma_{H\to gg}$ contains contributions with a $gq\bar{q}$-vertex.
On the one hand, these are the processes
$gH\to q\bar q$ and $H\to gq\bar q$ with a thermal gluon and a hard quark-antiquark pair, which give
rise to the term $\propto  \nf \tf \langle  T^{00}_g\rangle$ in (\ref{eq13}). On the other hand, there are the processes
$q H\to q g$ and $H\to g q\bar q$ with a thermal quark which contribute to the term proportional to
$\langle T^{00}_{fq}\rangle$. In addition, the virtual quark-loop correction to $H\to gg$ is also proportional to
$\langle T^{00}_{fq}\rangle$.

The physical picture behind obtaining (\ref{eq13}) from the trivial insertion of (\ref{eq8}) into (\ref{eq5})
is that the hierarchy $M_H \gg T$
allows for a separation of short and long distance physics. The Higgs gluon coupling $C_{Hg}$
describes physics which takes place on length and time scales much shorter than $1/T$
and which is therefore not affected by the presence of the QCD plasma. The
long-distance physics is given by the OPE of the bulk channel spectral function
whose temperature dependence is parametrized by the thermal expectation values of the
quark and gluon condensates, $\langle T^{00}_{fq}\rangle$ and  $\langle  T^{00}_g\rangle $, respectively.
To leading order in $\als$, these are given by the free (Stefan-Boltzmann) limits
\begin{eqnarray}
	\langle  T^{00}_g\rangle &=& \frac{\pi^2 T^4}{15} d_A\, , \label{eq14}\\
	\langle  T^{00}_{fq}\rangle \Big\vert_{m_q=0} &=& \frac{7 \pi^2 T^4}{60} d_F\, ,
	\label{eq15}
\end{eqnarray}
where $d_A = N_c^2 - 1$ and $d_F=N_c$ are the dimensions of the adjoint and the fundamental representation, respectively.
If a quark has mass $m_q\lesssim T$ one would need the explicit evaluation of the massive Stefan-Boltzmann integral instead of (\ref{eq15}),
while for $m_b \gg T$, $\langle T^{00}_{fb} \rangle$ is exponentially suppressed. The sum $\sum_{q\in udsc}$ in (\ref{eq13}) thus goes over the
flavors that can be thermally excited. To arrive at a more compact expression, one may approximate this sum by
an effective number  $\nf^T$ of approximately massless flavors, using $3<\nf^T<4$
for temperatures well above the strange quark mass and well below the bottom charm mass.
For the number of flavors entering the leading coefficient $b_0$ of the $\beta$-function, we use
$\nf \tf = \textstyle\frac{5}{2}$ in (\ref{eq13}).
  With this input, we obtain
\begin{eqnarray}
		&&\delta\Gamma_{H\to gg} = - \Gamma_{H\to gg}^\mathrm{vac} \als \frac{T^4}{M_H^4} \frac{112\, \pi^3}{45}
						\left(8-\nf^T\right)\, , \nonumber \\
						&& \qquad  \hbox{for $H$-decay in the plasma rest frame}\, .
	  \label{eq16}
\end{eqnarray}

\subsection{$H\to \bar{q}\, q$}
The decay of the Higgs boson into a $q\bar{q}$ pair proceeds via coupling to the scalar operator $S$.
For $m_q\ll M_H$, the leading  order
vacuum contribution to the corresponding spectral function $\rho_S$ is
\begin{equation}
	\rho^\mathrm{vac}_{S}(K^2)=\frac{d_F\nf m_q^2 K^2}{4\pi}\, ,
	\label{eq17}
\end{equation}
and its leading
(dimension-four) thermal correction reads~\cite{CaronHuot:2009ns}
\begin{eqnarray}
	\delta \rho_S(K) &=&
	  \frac{8\als m_q^2}{3K^2} \frac{K_\mu K_\nu}{K^2}
	      \left[\frac{13}{2}\cf T^{\mu\nu}_f - \nf\tf T^{\mu\nu}_g \right] \nonumber \\
	      &&	      -\frac{9\als m_q^2\cf}{K^2}\, S\, . 
	      \label{eq18}
\end{eqnarray}
Inserting the vacuum contribution  (\ref{eq17})  into (\ref{eq4}), we reproduce for each mass state ($\nf = 1$) the
LO vacuum branching ratio
\begin{equation}
	\label{eq19}
		\Gamma^\mathrm{vac}_{H\to q\bar q}
		=  \frac{d_F m_q^2 M_H}{8\pi v^2},
\end{equation}
which agrees with the literature~\cite{Djouadi:2005gi}.
(Full accounting of the massive kinematics amounts to a multiplicative factor $(1-4m_q^2/M_H^2)^{3/2}$.)

Having checked in this way the consistency of the normalization of $\rho_S$ and (\ref{eq4}), one can proceed to
determining in the same way the thermal correction to $\Gamma^\mathrm{vac}_{H\to q\bar q}$ from $\delta \rho_S(K)$
in eq.~(\ref{eq18}).  In general, the evaluation of the operator (\ref{eq18}) in the QCD plasma requires the LO thermal
(Stefan-Boltzmann) expectation value of the chiral condensate
\begin{equation}
	\langle S \rangle=4 d_F m_q^2 \int\frac{d^3p}{(2\pi)^3} \frac{ n_F(E_p)}{E_p}\, ,
	\label{eq20}
\end{equation}
which becomes
$\textstyle\frac{1}{6} d_F m_q^2 T^2$ for $m_q\ll T$.
However, for the thermal corrections to $H\to b\bar{b}$ at temperature $T\ll m_b$,
the contributions $\langle S \rangle$ and $\langle T^{00}_f \rangle$ in (\ref{eq18}) are
exponentially suppressed by the quark mass, and
\begin{equation}
	\label{eq21}
		\delta\Gamma_{H\to b\bar b}=-\frac{4\als\tf m_b^2}{v^2k_0}\frac{3k_0^2+k^2}{9M_H^4}
		 \langle T^{00}_g\rangle + \dots,
\end{equation}
where the dots stand for $ \OO\left(\exp\left[ -m_b/T\right]\right)$ terms.  Inserting the LO expression (\ref{eq14}) for the gluon condensate, we find (for $T\ll m_b$)
\begin{eqnarray}
		&&\delta\Gamma_{H\to b\bar{b}}= - \Gamma_{H\to b\bar{b}}^\mathrm{vac} \als \frac{T^4}{M_H^4} \frac{128\, \pi^3}{135} \, , \nonumber \\
			&& \qquad	 \hbox{for $H$-decay in the plasma rest frame}\, .
	  \label{eq22}
\end{eqnarray}
For temperatures $T \gtrsim \OO(m_b)$ or for the calculation of the partial thermal width into lighter quarks,
the contributions $\langle S \rangle$ and $\langle T^{00}_f \rangle$
in (\ref{eq18}) need to be included.
In general, the thermal corrections stemming from the coupling to lighter quarks are reduced by a factor
$m_q^2/m_b^2$ compared to (\ref{eq22}). For all partial decay widths into $q\bar{q}$-pairs, the thermal correction is an $\OO\left(\als \textstyle\frac{T^4}{M_H^4} \right)$
correction to the vacuum width.

\subsection{Thermal corrections to spectral functions: range of validity}
\label{sec2.3}
Here, we shortly recall the derivation of thermal corrections to $\rho_J(K)$ in the OPE approach~\cite{CaronHuot:2009ns}, and we comment on its range of validity.
The starting point is the Euclidean current-current correlator $G_E(q) = \int d^4x\, e^{-iq.x} \langle J(x)\, J(0)\rangle$, where we set $q = (0,0,0,q_E)$ for simplicity.
The dispersion relation $G_E(q_E) = P(q_E) + \int_{-\infty}^{+\infty} \textstyle\frac{d\omega}{2\pi (\omega-iq_E)} \rho_J(\omega)$
relates $G_E$ to the spectral function $\rho_J$, with $P(q_E) $ a polynomial in $q_E$.  As explained in detail in Ref.~\cite{CaronHuot:2009ns}, this dispersion relation
implies that the asymptotic expansion of $\rho_J(K)$
for large time-like $K$ can be obtained from matching term-by-term to the
operator product expansion of $G_E(q_E)$
for large space-like $q_E$,
\begin{eqnarray}
	&&G_E(q_E) \sim \sum_n \langle \OO_n\rangle \frac{c_n}{q_E^{d_n}} \nonumber \\
		&&  \Longleftrightarrow \nonumber \\
	&& \rho_J(k_0)\sim \sum_n \langle \OO_n\rangle 2 {\rm Im}\left[ \frac{c_n}{(-i k_0)^{d_n}} \right]\, .
	\label{eq23}
\end{eqnarray}
In practice, one proceeds as follows:
First, expand the operator product $J\, J$ in $G_E(q)$ up to order $1/q^2$ in the Euclidean four-momentum squared.  This results in explicit expressions such as
$G^\zeta_E(q) \sim 4\, b_0^2\als^2 \left( \textstyle\frac{q_\mu q_\nu}{q^2} \langle T_g^{\mu\nu}  \rangle + \textstyle\frac{1}{g^2} \langle \OO_{Hg} \rangle \right)$
for the bulk viscous channel.  Then take into account that the local operators in this expansion are scale dependent, for instance,
$T_g^{\mu\nu}(q_E) \sim T_g^{\mu\nu}(\mu_0) +
\textstyle\frac{\als}{3\pi} \log\left[\textstyle\frac{\mu_0^2}{q_E^2} \right] \left(\nf\tf T_g^{\mu\nu}(\mu_0) - 2\nf T_f^{\mu\nu}(\mu_0) \right)$.
This renormalization group flow is of central importance, since the branch cuts of the analytically continued logarithms
$\log\left[\textstyle\frac{\mu_0^2}{(-i K)^2} \right]$ contribute to ${\rm Im}G_E(-i K)$ and thus to $\rho_J$ at large time-like momenta.
Without this RG flow, the expansion of $G_E(q)$ would contain only powers of the type $1/q^n$ times local operators. The analytic continuation
of these $1/q^n$-terms to Minkowksi space can only generate discontinuities on the light cone. The only contribution to $\rho_J(K)$ at large time-like $K$
thus comes from these analytically continued logarithms. 

The OPE of $G_E$ in (\ref{eq23}) implements a physical scale separation. For a highly energetic, short-distance probe that tests distances of size $1/q_E$  much
smaller than any other scale in the problem, $1/q_E \ll 1/T$, eq.~(\ref{eq23}) systematically expands in powers of that small scale times local operators. For the corresponding
spectral function $\rho_J(K)$ to be valid, it is thus a necessary condition that
\begin{equation}
	K^2 = 4 k^+\, k^- = M^2 \gg T^2\, ,
	\label{eq24}
\end{equation}
where we have introduced the light-cone momenta $k^+ = \textstyle\frac{1}{2} \left( k_0 + k\right)$, $k^- = \textstyle\frac{1}{2} \left( k_0 - k\right)$.
In a thermal medium and for a very massive probe, $k^+ \gg T$ is always satisfied. However, to a boosted probe, the medium appears Lorentz-contracted, and the
scale separation between the long-distance physics of the medium and the short-distance physics of the probe becomes questionable when the coherence length $\sim 1/k^-$
of the probe becomes comparable to the medium scale $1/T$. One should therefore distinguish the following kinematic regimes:
\begin{enumerate}
	\item $k^- \gg T$:  $\rho_J$ can be determined from OPE.
	\item $k^- \sim T$: $\rho_J(K)$ cannot be determined from OPE, but unresummed perturbative techniques such as those used in Ref.~\cite{Laine:2011xm,Laine:2013vpa,Laine:2013lka,Laine:2013vma} apply  for `hard' momenta $k^- \sim \OO(T)$.
	\item $k^- \ll T$: Resummed finite temperature perturbation theory or non-perturbative methods would be needed to determine $\rho_J(K)$ in this regime,
	as in
\cite{Aurenche:2002wq,Ghisoiu:2014mha,Ghiglieri:2014kma}.

\end{enumerate}
For the Higgs boson decay discussed in this section, $k^- \gg T$ applies as long as the three-momentum $k$ in the medium satisfies $k \ll \textstyle\frac{M_H^2}{4\, T}$.
For temperature $T\leq 1$ GeV that may be reached in heavy ion collisions at present or future colliders, the OPE and the results for the partial thermal widths
(\ref{eq13}) and (\ref{eq21}) of the Higgs boson that we derived from it are thus valid over a transverse momentum range that extends to multiples of the Higgs mass.
Over this range of validity of the OPE, thermal corrections to $\Gamma_H$ are seen to increase by a factor $\textstyle\frac{3k_0^2 + k^2}{3 M_H^2}$ with the Higgs three-momentum $k$.
Finally, we note that the unresummed perturbative calculations of
spectral functions of the kind being considered here for
$M\gtrsim T$ find that the OPE regime sets in when $M$ is approximately
an order of magnitude larger than $T$ \cite{Laine:2011xm,Laine:2013vpa,Laine:2013lka,Laine:2013vma}. As $M_H$
is two orders of magnitude larger than the temperatures of QCD plasmas,
the applicability of the OPE expansion is thus certain.

 \section{Conclusions}
 For a Higgs boson at rest in a QGP of temperature $T \ll M_H$, explicit expressions for the thermal corrections to the partial decay widths $\Gamma^\mathrm{vac}_{H\to gg}$ and
 $\Gamma^\mathrm{vac}_{H\to q\bar{q}}$ are given in eqs. (\ref{eq16}) and (\ref{eq22}). These corrections are $ \OO\left( \als \left( \textstyle\frac{T}{M_H}\right)^4 \right)$ times the vacuum
 branching ratios.

 For a Higgs boson propagating with finite three-momentum $k$ through the QGP, the thermal width increases with $k$ like $\delta \Gamma^{k=0}\times\left(1 + \textstyle\frac{4}{3}  \textstyle\frac{k^2}{M_H^2}\right)$.
 This applies for $k^- = \textstyle\frac{1}{2} \left( k_0 - k\right) \gg T$, a range of validity which includes for temperatures $T< 1$ GeV even moderately relativistic Higgs bosons in the QGP.

 In general, the $ \OO\left( \als \left( \textstyle\frac{T}{M}\right)^4 \right)$
  leading thermal corrections to the decay width of neutral massive particles
	is caused by the
  absence of lower-dimension gauge-invariant local operators in QCD.
	For the thermal width of the Higgs,
 the $T^4$-dependence arises from the Stefan-Boltzmann limits
 of the quark (\ref{eq15}) and gluon (\ref{eq14}) condensates that enter thermal corrections of the spectral functions (\ref{eq8}) and (\ref{eq18}) of the bulk viscous and scalar operator, respectively.
 Similarly, thermal  corrections to the width of the $Z$-boson are $ \OO\left( \als \left( \textstyle\frac{T}{M_Z}\right)^4 \right)$, since the spectral functions of the vector and axial vector currents
 receive the dominant thermal corrections from the same quark and gluon condensates, see Ref.\cite{CaronHuot:2009ns} and Appendix~\ref{appb}.

We note that the leading $T$-dependence can be larger in
theories with lower-dimension gauge-invariant local operators. In the heavy
sterile neutrino case mentioned before, the zero-temperature
decay into a Higgs scalar and a SM lepton receives an $ \OO\left( \lambda \left( \textstyle\frac{T}{M}\right)^2 \right)$ correction
 \cite{Salvio:2011sf,Laine:2011pq,Biondini:2013xua}. This is due to
the dimension-two $\phi^\dagger\phi$ condensate of the Higgs field, with
 its  self-coupling $\lambda$.

\acknowledgements
 We thank D. d'Enterria, C. Loizides, P. Nason, G. Salam and G. Zanderighi for discussions at an early stage of this work.

\appendix
\section{Non-OPE results for the $H\to gg$ coupling at NLO}
\label{appa}
As mentioned in the introduction, explicit perturbative calculations of thermal widths proceed by calculating IR-regulated thermal corrections
to real emission, real absorption and virtual terms in the branching process. It is only the sum of these three contributions that is physically meaningful
and IR-safe. The present appendix provides technical details of how $\delta \rho_{\OO_{Hg}}(K)$, used to calculate $\delta \Gamma_{H\to g\, g}$ in (\ref{eq13}),
can be understood as arising from  the sum of these three IR-sensitive contribution. These details are not needed to follow our derivation of thermal widths.
We include them solely since they may help to understand the relation between the OPE approach followed here, and explicit perturbative calculations of thermal widths.
The following discussion is limited to the pure glue part of $\delta \rho_{\OO_{Hg}}(K)$, and to $k=0$.
It starts from the detailed calculation of the NLO bulk viscous spectral function, given in Ref.~\cite{Laine:2011xm} for pure Yang-Mills theory
for $k_0\gtrsim T$. Its applicability is hence wider than the $k_0\gg T$
region, and it provides a derivation of thermal corrections to the spectral function that is logically independent of the OPE and that verifies the results of the OPE.

Ref.~\cite{Laine:2011xm} calculates $\rho_{\OO_{Hg}}(K)$ at NLO in the imaginary-time formalism of thermal
perturbation theory. To this end, the contributions to the  $T^\mu{}_\mu\,T^\nu{}_\nu$ correlator are written without performing the sum integrals, and the sum of the amplitudes
is reduced to a set of master  two-loop amplitudes.
One then performs first the Matsubara sums, then one analytically continues the external Euclidean frequency $k_n$ to the Minkowskian $k_0+i\epsilon$, and one finally
takes the imaginary part to obtain the spectral function.

Taking this imaginary part corresponds to taking the sum over all possible cuts. At this stage, identifying the different real and virtual cut contributions to the spectral function
requires introducing an IR-regulator for the soft and collinear divergences in the cuts. Different regularization schemes
are possible. In Ref. \cite{Laine:2011xm}, the authors supplement one of the propagators in the master amplitudes with a regulating mass term $\lambda$.
Once the regulator has been introduced, each cut of each master amplitude is reduced to a set of one-  or two-dimensional integrals. Upon summing the cuts,
the $\lambda$-dependence disappears and the integrals are evaluated numerically. The final physical result is  scheme independent, and thus finite, for $\lambda\to 0$.

Here, we reverse-engineer the last step of this calculation. In the appendices A and B of Ref.~\cite{Laine:2011xm}, the real cuts are called ``phase-space integrals'' and the virtual
ones ``factorized integrals''. For each cut, they can be evaluated  after subtraction of the vacuum contribution for $k_0\gg T\gg\lambda$.  In this limit,  many terms become exponentially suppressed
($\exp(-k_0/T)\approx 0$). In particular, all two-dimensional
integrals are exponentially suppressed, and one has to deal only with the easier one-dimensional ones, which we
can integrate analytically for $k_0\gg T\gg\lambda$.

\begin{widetext}
For the virtual contribution, one obtains in this way in the scheme of \cite{Laine:2011xm} and taking the normalization of $\OO_{Hg}$ into account
\begin{align}
	\delta \rho_{\OO_{Hg}}(k_0,\lambda)\bigg\vert_\mathrm{virt}  =& \frac{d_A g^2 \nc}{32 \pi}\bigg\{
	-\frac{k_0^4}{2 \pi ^2} \left[\frac{2 \pi
		   T}{\lambda}-\ln ^2\left(\frac{\pi  T}{\lambda}\right)+2(\ln (4)- \gamma_E )\ln \left(\frac{\lambda}{4\pi  T}
		   \right)+2 \gamma _1-\frac{\pi ^2}{6}-\ln^2(4)\right]\nn\\
&-\frac{k_0^3
	   T}{8}	-\frac{k_0^2 T^2}{6}  \bigg[-144 \ln (A)
	+4 \ln \left(\frac{64 \pi^3  k_0 T^3}{\lambda^4}\right)
	+11
	\bigg]\nn\\
	&+\frac{16 T^4 }{45 \pi ^2}\left[\pi ^4 \left(-3 \ln \left(\frac{k_0 T}{\lambda^2}\right)+3
	   \gamma_E -5-\ln (8)\right)-270 \zeta '(4)\right] + \OO\left( \frac{T^5}{k_0}\right)
	   \bigg\}\, ,
	   \label{mikkovirtual}
\end{align}
where $\gamma_E$ is the Euler-Mascheroni constant, $\gamma_1$ is the first Stieltjes constant and
$\ln(A)=1/12-\zeta'(-1)$ is the logarithm of Glaisher's constant.

The real emission contribution ($H\to ggg$) is instead
\begin{align}
	\delta \rho_{\OO_{Hg}}(k_0,\lambda)\bigg\vert_\mathrm{emi}  =& \frac{d_A g^2 \nc}{32 \pi}\bigg\{
	\frac{k_0^4}{4 \pi ^2} \left[\frac{2 \pi
		   T}{\lambda}-\ln ^2\left(\frac{\pi  T}{\lambda}\right)+2(\ln (4)- \gamma_E )\ln \left(\frac{\lambda}{4\pi  T}
		   \right)+2 \gamma _1-\frac{\pi ^2}{6}-\ln^2(4)\right]\nn\\
	   	&+\frac{k_0^3 T}{8 \pi ^2} \left[ \ln ^2\left(\frac{2 k_0 T}{\lambda^2}\right)-10 \ln
	   	   \left(\frac{T}{\lambda}\right) \ln \left(\frac{4 T}{\lambda}\right)+18 \gamma _1-\frac{\pi ^2}{3}+9 \gamma_E
	   	   ^2-10 \ln ^2(2)\right]\nn\\
	&+\frac{k_0^2 T^2}{12}  \bigg[-144 \ln (A)
	+4 \ln\left(\frac{64\pi^3k_0T^3}{\lambda^4}\right)
	+11\bigg]\nn\\
   &+\frac{k_0 T^3}{\pi ^2} \left[\zeta (3) \left(\ln
   \left(\frac{k_0^4}{16 T^4}\right)+4 \gamma_E -15\right)-4 \zeta '(3)\right]\nn\\
	   &+\frac{8 T^4}{45 \pi
	   ^2} \left[\pi ^4 \left(3 \ln
	   \left(\frac{k_0 T}{\lambda^2}\right)-3 \gamma_E -\frac12+\ln (8)\right)+270 \zeta '(4)\right] + \OO\left( \frac{T^5}{k_0}\right)
	   \bigg\}\, .
	   \label{mikkoemi}
\end{align}
Finally, the absorption contribution ($gH\to gg$) reads
\begin{align}
	\delta \rho_{\OO_{Hg}}(k_0,\lambda)\bigg\vert_\mathrm{abs}  =& \frac{d_A g^2 \nc}{32 \pi}\bigg\{
	\frac{k_0^4}{4 \pi ^2} \left[\frac{2 \pi
		   T}{\lambda}-\ln ^2\left(\frac{\pi  T}{\lambda}\right)+2(\ln (4)- \gamma_E )\ln \left(\frac{\lambda}{4\pi  T}
		   \right)+2 \gamma _1-\frac{\pi ^2}{6}-\ln^2(4)\right]\nn\\
	   	&-\frac{k_0^3 T}{8 \pi ^2} \left[ \ln ^2\left(\frac{2 k_0 T}{\lambda^2}\right)-10 \ln
	   	   \left(\frac{T}{\lambda}\right) \ln \left(\frac{4 T}{\lambda}\right)+18 \gamma _1-\frac{4\pi ^2}{3}
		   +9 \gamma_E
	   	   ^2-10 \ln ^2(2)\right]\nn\\
	&+\frac{k_0^2 T^2}{12}  \bigg[-144 \ln (A)
	+4 \ln\left(\frac{64\pi^3k_0T^3}{\lambda^4}\right)
	+11\bigg]\nn\\
   &-\frac{k_0 T^3}{\pi ^2} \left[\zeta (3) \left(\ln
   \left(\frac{k_0^4}{16 T^4}\right)+4 \gamma_E -15\right)-4 \zeta '(3)\right]\nn\\
	   &+\frac{8 T^4}{45 \pi
	   ^2} \left[\pi ^4 \left(3 \ln
	   \left(\frac{k_0 T}{\lambda^2}\right)-3 \gamma_E -\frac12+\ln (8)\right)+270 \zeta '(4)\right] + \OO\left( \frac{T^5}{k_0}\right)
	   \bigg\}.
	   \label{mikkoabs}
\end{align}
\end{widetext}
Upon summing the three contributions all divergent terms, as well as all terms larger than $\OO(T^4)$, cancel out,
yielding
\begin{eqnarray}
	\delta \rho_{\OO_{Hg}}(k_0)\bigg\vert_\mathrm{tot} &=&
	\delta \rho_{\OO_{Hg}}(k_0,\lambda)\bigg\vert_\mathrm{virt} +
	\delta \rho_{\OO_{Hg}}(k_0,\lambda)\bigg\vert_\mathrm{emi} \nonumber \\
	&& +
	\delta \rho_{\OO_{Hg}}(k_0,\lambda)\bigg\vert_\mathrm{abs} \nonumber \\
	&=& -\frac{11 \pi^2 d_A \nc\als  T^4}{45}.
	\label{totmikko}
\end{eqnarray}
If we take Eq.~\eqref{eq8} and set $\nf=0$, $k=0$ we have
\begin{equation}
	\label{oTmikko}
	\delta \rho_{\OO_{Hg}}(k_0)\bigg\vert_{\nf=0}  =
	 - \als
	   b_0 \langle T^{00}_g \rangle= -
	 \frac{ 11\pi^2 d_A\nc\als T^4}{45},
\end{equation}
which agrees as expected. The material in this apendix further illustrates the complexity of perturbative calculations compared to the
relative simplicity of deducing thermal corrections to the width from the OPE approach. The three contributions \eqref{mikkovirtual}, \eqref{mikkoemi} and \eqref{mikkoabs}
depend, of course, on the IR regularization scheme. They illustrate, however, how the different scheme-dependent IR-singular cut contributions in a perturbative
calculation sum up to a physical result that is free of any IR regulator.

\section{$Z$-boson thermal widths}
\label{appb}
In between the lines of Ref.~\cite{CaronHuot:2009ns}, one reads that it was one motivation for Caron-Huot's study of the asymptotic behavior of spectral
function to clarify in a logically independent way the $\textstyle\left(\frac{T}{M_Z}\right)$-dependence of the thermal width $\delta \Gamma_Z$ of the $Z$-boson
for which different explicit perturbative calculations had obtained different power laws. However, despite this motivation, and despite stating clearly that the
leading thermal correction in this case is $ \OO\left( \als \left( \textstyle\frac{T}{M_Z}\right)^4 \right)$, the results of Ref.~\cite{CaronHuot:2009ns}
have never been used to write an explicit expression for  $\delta \Gamma_Z$. This appendix aims at filling this small gap in the existing literature.

The $Z$-boson decay to $q\bar{q}$-pairs is mediated by coupling to the vector and axial vector currents
\begin{eqnarray}
	\label{eqB1}
	\Gamma_{Z\to q\bar q} &=&\frac{g_1^2+g_2^2}{6 k^0}\left(-g_{\mu\nu}+\frac{k_\mu k_\nu}{M_Z^2}\right) \nonumber \\
	&& \bigg[g_V^2\,\rho_V^{\mu\nu}(K)+g_A^2\rho_A^{\mu\nu}(K)\bigg]+\OO(\als)\, .
\end{eqnarray}
Here, the vector and axial vector spectral functions $\rho_V$ and $\rho_A$ couple with
$g_V=1/2 T_3-Q\sin^2\theta_W$, $g_A=T_3/2$, respectively, where $T_3=\pm1/2$ for up/down-type quarks and $Q=+2/3$
for up-type, $Q=-1/3$ for down type. The factor of $1/(6k^0)$ is a combination of the usual flux factor $1/(2k^0)$
times the average over the 3 polarization states of the $Z$ boson.

For the conserved vector current we can assume $k$ to point in
the $z$ direction and define $\rho_T\equiv\rho^{xx}=\rho^{yy}$,
$\rho_L\equiv \textstyle\frac{K^2}{k_0^2} \rho^{zz}=  \textstyle\frac{K^2}{k^2}\rho^{00}$. Hence
\begin{eqnarray}
	\label{eqB2}
	\Gamma_{Z\to q\bar q} &=& \frac{g_1^2+g_2^2}{6 k^0}
	\bigg[g_V^2\,\big(2\rho_T(K)+\rho_L(K)\big) \nonumber \\
	&& +g_A^2\bigg(2\rho_A^{xx}(K)+\frac{k_0^2}{M_z^2}\rho_A^{zz}(K)
	+\frac{k^2}{M_z^2}\rho_A^{00}(K) \nonumber \\
	&& \qquad \qquad -2\frac{k^0k}{M_z^2}\rho_A^{0z}(K)\bigg)\bigg]+\OO(\als),
\end{eqnarray}
At vanishing quark mass the axial vector current becomes also (classically) conserved, so that
we can use
\begin{equation}
	\rho_T^\mathrm{vac}(K)=\rho_L^\mathrm{vac}(K)=\frac{\nf d_F K^2}{6\pi}+\OO(\als)
\end{equation}
for vector and axial current alike.
This yields
\begin{equation}
	\label{vacZfinal}
	\Gamma^\mathrm{vac}_{Z\to q\bar q}=\frac{(g_1^2+g_2^2)\nf d_F M_Z^2}{12 \pi k^0}
	\big(g_V^2\,+g_A^2\big)+\OO(\als)\, ,
\end{equation}
which is a limit of the well-known expression for a non-negligible mass
\begin{eqnarray}
	\label{vacZfinalmass}
	\Gamma^\mathrm{vac}_{Z\to q\bar q}\! &=&\! \frac{(g_1^2{+}g_2^2)d_F}{12\pi k^0}
	\bigg[\big(g_V^2{+}g_A^2\big)M_Z^2+2(g_V^2-2g_A^2)m_q^2\bigg] \nonumber \\
	&&	\sqrt{1-\frac{4m_q^2}{M_Z^2}}+\OO(\als)\, ,
\end{eqnarray}

For the thermal width, we need the thermal corrections to the longitudinal and transverse pieces of the
vector current~\cite{CaronHuot:2009ns}
\begin{align}
	\delta \rho_T(K)  &=
	\frac{16\als}{9K^2}
	 \frac{k_0^2{+}k^2}{K^2}
	      \left[2\cf T^{00}_f {-} \nf\tf T^{00}_g \right]\, ,
	 \label{transT}
	\\
	\delta \rho_L(K)  &=
	 \frac{16\als}{9K^2}
	      \left[2\cf T^{00}_f {-} \nf\tf T^{00}_g \right]\, .
	 \label{longT}
\end{align}

To also obtain the corresponding thermal correction to the spectral function of the axial vector current, one can parallel for $J_A$ the analysis of the Euclidean $J_V\, J_V$
operator product  in eqs.~(3.3) and (3.4) of Ref.~\cite{CaronHuot:2009ns}. One finds that up to dimension four, this OPE is expressed in terms of
two local operators $T^{44}_f$ and $\OO_m$, where the index $4$ denotes the Euclidean time. Since $T^{44}_f$ is the operator that survives in the chiral limit,
it has the same Wilson coefficient in the OPE of $J_A\, J_A$ and $J_V\, J_V$. The operator  $\OO_m$  occurs with different Wilson coefficients in both current products, and the
terms that violate current conservation are found to be proportional to it. But  $\OO_m$ does not matter because
it is RGE invariant and does not generate cuts, so the thermal corrections to $\rho_A$ have to agree with
those of $\rho_V$ even at nonzero $m_q$ (as long as $m_q\ll M_Z$). Hence the thermal correction
to the $Z$ width into a quark of a particular flavor ($n_f = 1$) can be written as
\begin{eqnarray}
	\label{ZT}
	\delta\Gamma_{Z\to q\bar q} &=&\frac{g_1^2+g_2^2}{3 k^0}\frac{8\als(3k_0^2+k^2)}{9M_Z^4}
	\big(g_{Vq}^2+g_{Aq}^2\big)  \\
	&&	\bigg( 2\cf \langle T^{00}_{fq}\rangle - \tf \langle T^{00}_g\rangle
	 \bigg)+\OO(\als^2)
	 \nonumber \\
	 &=& \Gamma^\mathrm{vac}_{Z\to q\bar q} \als \frac{32\, \pi}{27} \frac{3k_0^2 + k^2}{M_Z^2} \frac{2\cf \langle T^{00}_{fq}\rangle - \tf \langle T^{00}_g\rangle }{M_Z^4}\, .
	 \nonumber
\end{eqnarray}
For the light $uds$ quarks in the QGP one can assume $m_q=0$ and take the massless expressions in eqs.~\eqref{eq14} and ~\eqref{eq15}.
One then finds $2\cf \langle T^{00}_{fq}\rangle - \tf \langle T^{00}_g\rangle = \textstyle\frac{2}{3} \pi^2 T^4$, which yields e.g.
\begin{equation}
	\delta\Gamma_{Z\to u\bar u} =   \Gamma^\mathrm{vac}_{Z\to q\bar q}
	\als \frac{64\, \pi^3}{81} \frac{3k_0^2 + k^2}{M_Z^2} \frac{T^4}{M_Z^4}\, ,
	\label{ZTlight}
\end{equation}
and identical for the branching into $d$- and $s$-quarks. The thermal correction is again an effect of $ \OO\left( \als \left( \textstyle\frac{T}{M_Z}\right)^4 \right)$, but in contrast to
the standard model Higgs boson, it comes for light quarks with a positive sign.

We note that for the closely related case of dilepton production at $\frac{T}{M_{l\bar l} } \ll 1$, the correct $ \OO\left( \als \left( \textstyle\frac{T}{M_{l\bar l} }\right)^4 \right)$ was found
already in Ref.~\cite{Altherr:1989jc}.

For the branching of the $Z$-boson into $b \bar b$ or $c \bar c$ quark pairs and for temperatures relevant for heavy-ion collision experiments,
the fermion condensate in \eqref{ZT} should be evaluated for massive quarks, i.e.,
\begin{equation}
	\label{massiveSB}
	\langle T^{00}_f\rangle=4 d_F \int\frac{d^3p}{(2\pi)^3} E_p n_F(E_p)
	- d_F \int\frac{d^3p}{(2\pi)^3} \frac{m^2}{E_p} n_F(E_p)\, .
\end{equation}
For a sufficiently large ratio of quark mass over temperature, the contribution $\langle T^{00}_{fQ}\rangle$ in \eqref{ZT}  becomes exponentially suppressed and can be neglected.
While the suppression factor $\propto \als \left( \textstyle\frac{T}{M_Z}\right)^4$ will render all these effects unobservable in practice, it is still curious to note that the thermal correction
$\delta\Gamma_{Z\to Q\bar Q}$ to sufficiently heavy quarks will be dominated by the gluon condensate and therefore have a negative sign, in contrast to \eqref{ZTlight}.

\end{document}